\documentclass[aps,prd,twocolumn]{revtex4}
\usepackage{amssymb,amsmath,bm,graphicx,thmsa}
\begin{document}

\title{Estimates for first-order homogeneous linear characteristic problems}

\author{Simonetta Frittelli}
\email[]{simo@mayu.physics.duq.edu}
\affiliation{Department of Physics, Duquesne University,
        Pittsburgh, PA 15282}
\affiliation{Department of Physics and Astronomy, University of Pittsburgh,
       Pittsburgh, PA 15260}

\date{\today}

\begin{abstract}

An algebraic criterion that is sufficient to establish the existence of certain
\textit{a priori} estimates for the solution of first-order homogeneous linear
characteristic problems is derived. Estimates of such kind ensure the stability
of the solutions under small variations of the data. Characteristic problems
that satisfy this criterion are, in a sense, \textit{manifestly well posed.}

\end{abstract}
\maketitle

\section{Introduction}
\label{sec:1}

Any system of hyperbolic partial differential equations can be written in a
peculiarly wicked form, namely: in characteristic form~\cite{courant}.  In
order to do this, a one-parameter family of characteristic surfaces is chosen
as the level surfaces of a coordinate $u$, referred to as the null coordinate
or \textit{retarded time\/}. Instead of evolution from one time-level to the
next, one obtains evolution from one characteristic slice to the next. However,
characteristic surfaces are special: they are the only data surfaces for which
the standard Cauchy problem can not be solved, because the differential
operator of the system is internal to these surfaces, failing to evolve some
data out of the surface. Naturally, the equations ``degenerate'', turning into
a system where not all the equations can be integrated forward in retarded
time, but there are ``rules'' that allow one to solve the problem in a
hierarchical manner, zig-zagging back and forth between the
equations~\cite{duff}.  These hierarchical rules are the hallmark of
characteristic evolution, making it significantly different from an initial
value problem. 

It has been known for quite some time~\cite{duff} that the characteristic
``Cauchy problem'' -- to obtain a unique solution from data given on a
characteristic surface-- can be solved so long as the data are split into two
separate sets: some data are given on the initial null slice, and the rest of
the data are given on another slice that must be transverse to the initial null
slice.  An issue that has attracted less attention is: if the data are
perturbed slightly, under what circumstances is the variation of the solution
under control?  Equivalently, will almost-zero data evolve into a solution that
is also close to zero?  We set up a framework in which to address this question
by defining certain types of estimates of the solution in terms of the free
data, after \cite{cmpwave}. Subsequently we derive an algebraic criterion that
is sufficient to determine whether the solutions satisfy such an \textit{a
priori} estimate, thus establishing their stability with respect to small
variations of the free data. This kind of stability is of relevance to
numerical applications. A prominent instance of the use of the characteristic
problem for numerical applications is that of the simulation of gravitational
waves by numerically integrating the characteristic formulation of the Einstein
equations~\cite{jefflivrev}. 

As argued extensively in \cite{cmpwave}, characteristic problems for which the
estimate can be established may be considered to be \textit{well posed\/} in
the sense that for each set of data the solution exists, is unique and depends
continuously on the free data.  In addition, characteristic problems that
satisfy the algebraic criterion developed here can be thought of as
\textit{manifestly well posed}. Manifest well-posedness in the sense defined
here is to characteristic problems as symmetric hyperbolicity~\cite{kreissbook}
is to initial-value problems.  

Section~\ref{sec:2} describes
first-order linear characteristic problems after Duff~\cite{duff}. The estimates
of interest are defined in Section~\ref{sec:3} where the algebraic criterion is
developed as well. Concluding remarks are offered in Section~\ref{sec:4}. 

\section{Homogeneous linear characteristic
problems in canonical form}
\label{sec:2}

Consider a generic homogeneous hyperbolic system of linear partial
differential equations for $m$ functions $v~=~(v^\alpha)$ of $n$ variables
$y^a$, which can be written in matrix form as follows
\begin{equation}\label{eq:1}
	\bm{A^a}\frac{\partial v}{\partial y^a}+\bm{D}v = 0,
\end{equation}
 
\noindent where summation over repeated indices is understood. A
characteristic surface ${\cal N}$ is a surface given by $\phi(y^a)=0$
such that 
\begin{equation}\label{vanishdet}
	\det \left(\bm{A^a} \frac{\partial\phi}{\partial y^a}\right)=0.
\end{equation}

Denote by $m$ the multiplicity of this characteristic surface (so that
the rank of $\bm{A^a}\partial\phi/\partial y^a$ is $n-m$). Suppose ${\cal
T}$ given by $\psi(y^a)=0$ is another surface intersecting ${\cal N}$ at
a submanifold of dimension $n-2$, whose further properties are to be
determined.  We choose a suitable coordinate system $(u,x,x^i)$
$i=1\ldots n-2$ for $\mathbb{R}^n$ adapted to these two surfaces; i.e.,
such that
\begin{subequations}
\begin{eqnarray}
	u&\equiv&\phi(y^a),\\
	x&\equiv& \psi(y^a).
\end{eqnarray}
\end{subequations}

\noindent In these coordinates (\ref{eq:1}) reads
\begin{equation}\label{characsystem}
	\bm{B^u}\partial_u v+\bm{B^x}\partial_x v 
+\bm{B^i}\partial_i v +\bm{D}v = 0,
\end{equation}

\noindent with
\begin{subequations}
\begin{eqnarray}
	 \bm{B^u}&\equiv &\bm{A^a}\frac{\partial\phi}{\partial y^a},\\
	 \bm{B^x}&\equiv &\bm{A^a}\frac{\partial\psi}{\partial y^a},\\
	 \bm{B^i}&\equiv &\bm{A^a}\frac{\partial x^i}{\partial y^a}.
\end{eqnarray}
\end{subequations}

\noindent By (\ref{vanishdet}), there are $m$ linearly independent
left null vectors $\tilde{z}_{(\nu)}$ and also $m$ linearly
independent right null vectors $z_{(\nu)}$ (with $\nu=1\ldots m$) of
the matrix $\bm{B^u}$, namely
\begin{subequations}
\begin{eqnarray}
	\tilde{z}_{(\nu)}\bm{B^u}&=&0,\\
	\bm{B^u}z_{(\nu)}        &=&0.
\end{eqnarray}
\end{subequations}

\noindent We choose the right null vectors to be
orthonormal in the sense that 
\begin{equation}
	z_{(\nu)}^\alpha z_{(\mu)}^\alpha 
	= \delta_{\mu\nu}.
\end{equation}

\noindent Multiplying (\ref{characsystem}) on the left with 
$\tilde{z}_{(\nu)}$ we find that $m$ of the equations in the system
do not involve derivatives with respect to $u$:
\begin{equation}\label{hypersurface}
 \tilde{z}_{(\nu)}\bm{B^x}\partial_x v 
+\tilde{z}_{(\nu)}\bm{B^i}\partial_i v 
+\tilde{z}_{(\nu)}\bm{D}v = 0.
\end{equation}

\noindent Our aim is now to find a convenient transformation of
variables that takes advantage of this split of the equations.  We
start by noticing that, using the $m$ right null vectors as the first
$m$ legs of an orthonormal basis $e'_\alpha$ of $\mathbb{R}^m$, we
have a unitary transformation
\begin{equation}
	e'_\alpha = S_{\alpha\beta}e_\beta
\end{equation} 

\noindent from the trivial basis $e_\beta= \{(1,0,\ldots,0), \ldots, 
(0,\ldots,0,1)\}$
to the new orthonormal basis, with
$S_{\alpha\gamma}S_{\beta\gamma}=\delta_{\alpha\beta}$ and such that
$S_{\nu\alpha}= z_{(\nu)\alpha}$ for $\nu=1\ldots m$. The components
of $v=v_\alpha e_\alpha$ in the new orthonormal basis are
\begin{equation}\label{v'}
	{v'}_\alpha = S_{\alpha\beta}v_\beta.
\end{equation}

\noindent In particular, the first $m$ components are the scalar
products of $v$ with the right null vectors $z_{(\mu)}$, which we
denote by $w_{\mu}$
\begin{equation}
 w_\mu \equiv z_{(\mu)\alpha} v_\alpha 
= v'_\mu \hspace{0.5cm}\mbox{ for } \mu=1 \ldots m .
\end{equation}

\noindent Multiplying (\ref{characsystem}) on the left with $\bm{S}$,
the system transforms into
\begin{equation}\label{newcharac}
	\bm{{B'}^a}\partial_a v'+
\bm{D'}v' = 0.
\end{equation}

\noindent From now on the index $a$ refers to the characteristic 
coordinates, namely: $a=u,x,i$. The matrices have transformed according to  
$\bm{{B'}^a}=\bm{S}\bm{B^a}\bm{S^t}$ and 
$\bm{D'}=\bm{S}\bm{D}\bm{S^t}$, and $\bm{S^t}$ is the transpose of 
$\bm{S}$. Because  $(\bm{B^u}\bm{S^t})_{\alpha\nu} =
B^u_{\alpha\beta}S_{\nu\beta}= B^u_{\alpha\beta}z_{(\nu)\beta}=0$ for
$\nu\le m$, the matrix $\bm{{B'}^u}$ has a Jordan form with all
vanishing coefficients in the first $m$ columns.  This means that the
$u$-derivatives of the $m$ new variables $w_\nu$ are not involved,
and consequently, the remaing variables $v'_\nu$ with $\nu>m$ are the
normal variables of the problem, which we denote by $q=(q_\mu),
\mu=m+1\ldots n$. We have thus split the new fundamental
variables into 
\begin{equation}
	v' = \bm{S}v
	  \equiv (w_1\ldots w_m,q_{m+1},\ldots,q_{n}).
\end{equation}

\noindent Inverting (\ref{v'}) we have $v_\alpha =
S_{\beta\alpha}v'_\beta$, which can be used into (\ref{hypersurface})
to obtain a set of equations in the transformed variables:
\begin{equation}\label{newhyper}
\tilde{z}_{(\nu)}\bm{B^x}\bm{S^t}\partial_x v' 
+\tilde{z}_{(\nu)}\bm{B^i}\bm{S^t}\partial_i v' 
+\tilde{z}_{(\nu)}\bm{D}\bm{S^t}v' = 0.
\end{equation}	

\noindent We'd like for these equations to be solvable for the
$x$-derivatives of all the variables $w_\mu$, that
is: the ones that do not evolve out of the initial characteristic
surface. The first $m$ terms in each equation for fixed $\nu$ are 
\begin{equation}
\tilde{z}_{(\nu)\alpha}(\bm{B^x}\bm{S^t})_{\alpha\mu}\partial_x w_\mu
=\tilde{z}_{(\nu)\alpha}B^x_{\alpha\beta}z_{(\mu)\beta}\partial_x w_\mu
\end{equation}

\noindent Thus the set of $m$ equations (\ref{newhyper}) can be solved
for the $m$ variables $\partial_xw_\mu$ if and only if 
\begin{equation}\label{restriction}
\det\left(
\tilde{z}_{(\nu)\alpha}B^x_{\alpha\beta}z_{(\mu)\beta}
\right)\neq 0
\end{equation}

\noindent This is a restriction on the choice of $\psi(y^a)$. For this
restriction to hold it is sufficient, but not necessary, that the level
surfaces of $\psi(y^a)$ be non-characteristic. In many applications,
the level surfaces of $\psi$ are chosen to be timelike. For now on we
assume that (\ref{restriction}) holds.  This allows us to interpret the
$m$ variables $w_\nu$ as the null variables of the problem. 

We have shown that under very weak conditions for the surface ${\cal
T}$, the most general characteristic problem takes the following form
\begin{subequations}
\begin{eqnarray}
   \bm{N^u}\partial_u q
  +\bm{N^x}\partial_x q
  +\bm{N^i}\partial_i v'
  +\bm{N^0} v'
 &=& 0 \label{eq:pre-evolution}\\
   \partial_x w 
  +\bm{L^x}\partial_x q 
  +\bm{L^i}\partial_i v'
  +\bm{L^0} v'
  &=& 0	\label{eq:pre-hypersurface}
\end{eqnarray} 
\end{subequations}

\noindent Clearly the null variables $w$ can be redefined by
$\widehat{w} \equiv w+\bm{L^x}q$ so that none of the 
Eqs.~(\ref{eq:pre-hypersurface}) contains $x-$derivatives of the normal
variables. Additionally, since $\bm{N^u}$ is non-singular, we can
choose normal variables $\widehat{q}\equiv \bm{N^u}q$. In terms of
these special choices of null and normal variables, 
Eqs.~(\ref{eq:pre-evolution}) and (\ref{eq:pre-hypersurface}) assume what
is referred to as the \textit{canonical form}:
\begin{subequations}\label{eq:canonical}
\begin{eqnarray}
   \partial_u \widehat{q}
  +\bm{\widehat{N}^x}\partial_x \widehat{q}
  +\bm{\widehat{N}^i}\partial_i \widehat{v}
  +\bm{\widehat{N}^0}\widehat{v}
 &=& 0 \label{eq:evolution}\\
   \partial_x \widehat{w} 
  +\bm{\widehat{L}^i}\partial_i \widehat{v}
  +\bm{\widehat{L}^0} \widehat{v}
  &=& 0	\label{eq:hypersurface}
\end{eqnarray} 
\end{subequations}

\noindent where $\widehat{v} \equiv (\widehat{w},\widehat{q})$. We
refer to (\ref{eq:evolution}) as the \textit{evolution equations}, and
to (\ref{eq:hypersurface}) as the \textit{hypersurface equations}. For
a unique solution to exist, one must prescribe the values of
$\widehat{w}$ on the surface $x=0$ and the values of $\widehat{q}$ on
the surface $u=0$. The solution can then be constructed in a
hierarchical manner. Since $q$ is a known source for
(\ref{eq:hypersurface}) at $u=0$, then $\widehat{w}$ can be found on
the entire surface $u=0$.  Once $\widehat{w}$ is known at $u=0$, it can
be used as a given source for (\ref{eq:evolution}) in order to find the
values of the normal variables $\widehat{q}$ on the next surface at
$u=du$. These are then used into (\ref{eq:hypersurface}) to obtain
$\widehat{w}$ on the surface $u=du$. And so forth. In fact, Duff proves
a theorem of existence and uniqueness of the solution given the
canonical form of the characteristic problem~\cite{duff}.  

As an example, consider the following equations for four unknowns $v$
as functions of four variables $x^a=(t,x,y,z)$:
\begin{subequations}\label{cauchywave}
\begin{eqnarray}
      \partial_t v_1
  &=& \partial_x v_2
     +\partial_y v_3
     +\partial_z v_4 ,\\
      \partial_t v_2
  &=& \partial_x v_1 ,\\
      \partial_t v_3
  &=& \partial_y v_1 ,\\
      \partial_t v_4
  &=& \partial_z v_1  .
\end{eqnarray}
\end{subequations}

\noindent These equations constitute a first-order version of the
wave equation in three spatial dimensions (if we intepret the
variables $v_\alpha$ as the derivatives of a single function $f$).
However, this first-order version of the wave equation has
characteristic speeds of 0 (rest) in addition to 1 (light). The level
surfaces of  $\phi\equiv t-x$ are null planes, so they are
characteristic and intersect the surfaces of fixed value of $x$. We
change coordinates $(t,x,y,z)\to (u,x,y,z)$ with  
\begin{equation}
	u=t-x
\end{equation}

\noindent which implies that $\partial_t \to \partial_u$ and
$\partial_x \to\partial_x-\partial_u$. The system (\ref{cauchywave})
turns into  
\begin{subequations}\label{precharacwave}
\begin{eqnarray}
      \partial_u v^1
     +\partial_u v^2
  &=& \partial_x v^2
     +\partial_y v^3
     +\partial_z v^4 ,\\
      \partial_u v^2
     +\partial_u v^1
  &=& \partial_x v^1 ,\\
      \partial_u v^3
  &=& \partial_y v^1 ,\\
      \partial_u v^4
  &=& \partial_z v^1  .
\end{eqnarray}
\end{subequations}

\noindent We can read off the matrix $\bm{B^u}$:
\begin{equation}
\bm{B^u}=\left(\begin{array}{rrrr}
	1&1&0&0\\
	1&1&0&0\\
	0&0&1&0\\
	0&0&0&1
	\end{array}
\right)
\end{equation}

\noindent which is obviously singular of rank 3, so we have $m=1$ in
this example. We expect only one null variable, and three normal
variables for this problem in canonical form.  $\bm{B^u}$ has only
one right null vector $z=2^{-1/2}(1,-1,0,0)$, and only one left null
vector $\tilde{z}=2^{-1/2}(1,-1,0,0)$, which coincides with $z$
because $\bm{B^u}$ is symmetric. An orthonormal basis for
$\mathbb{R}^4$ can be chosen as
$\{2^{-1/2}(1,-1,0,0),2^{-1/2}(1,1,0,0), (0,0,1,0),(0,0,0,1)\}$. So
the unitary transformation is 
\begin{equation}
\bm{S}=\left(\begin{array}{rrrr}
	\frac{1}{\sqrt{2}}&-\frac{1}{\sqrt{2}}&0&0\\
	\frac{1}{\sqrt{2}}&\frac{1}{\sqrt{2}}&0&0\\
	0&0&1&0\\
	0&0&0&1
	\end{array}
\right)
\end{equation}

\noindent The null variable of the problem is 
$w=(v_1-v_2)/\sqrt{2}$ and the normal variables are
$q_\nu=((v_1+v_2)/\sqrt{2},v_3,v_4) $, in terms of which the system 
(\ref{precharacwave}) takes the almost canonical form:
\begin{subequations}\label{characwave}
\begin{eqnarray}
      2\partial_u q_2
     -\partial_x q_2
     -\frac{1}{\sqrt{2}}\partial_y q_3
     -\frac{1}{\sqrt{2}}\partial_z q_4
&=&0,\\
      \partial_u q_3
    -\frac{1}{\sqrt{2}}\partial_y q_2 
     -\frac{1}{\sqrt{2}}\partial_y w 
&=&0,\\
      \partial_u q_4
    -\frac{1}{\sqrt{2}}\partial_z q_2
     -\frac{1}{\sqrt{2}}\partial_z w
&=&0,\\  
      \partial_x w
   -\frac{1}{\sqrt{2}}\partial_y q_3
      -\frac{1}{\sqrt{2}}\partial_z q_4
&=&0.
\end{eqnarray}
\end{subequations}

\noindent For a unique solution, we need to prescribe the value of
$w$ on the surface $x=0$, and the values of $q_2,q_3$ and $q_4$ on
the surface $u=0$.  Notice that, in this example, the surface $x=0$
is not timelike with respect to the hyperbolic operator $\bm{A^a}$,
but is also characteristic, as can be seen by inspection of the
matrix $\bm{B^x}$:
\begin{equation}
\bm{B^x}=\left(\begin{array}{rrrr}
	0&-1&0&0\\
	-1&0&0&0\\
	0&0&0&0\\
	0&0&0&0
	\end{array}
\right)
\end{equation}

\noindent However, we have $\tilde{z}\bm{B^x}z=2\neq 0$. Therefore
the condition (\ref{restriction}) is satisfied in spite of the fact
that the surfaces of fixed value of $x$ are not timelike.

\section{Well-posedness of homogeneous linear characteristic problems}
\label{sec:3}

The canonical system (\ref{eq:canonical}) can be written in the 
compact form
\begin{equation}\label{eq:wellstart}
	\bm{C^a}\partial_a v + \bm{D}v =0
\end{equation}

\noindent where $\bm{C^u}$ and $\bm{C^x}$ have block-diagonal 
forms of a special type:
\begin{equation}
\bm{C^u}=\left(\begin{array}{cc}
	\bm{1}&\bm{0}\\
	\bm{0}&\bm{0}
	\end{array}
\right), 
\hspace{1cm}
\bm{C^x}=\left(\begin{array}{cc}
	\bm{N^x}&\bm{0}\\
	\bm{0}&\bm{1}
	\end{array}
\right) ,
\end{equation}

\noindent where $\bm{1}$ is the identity of dimension $n-m$ in the case
of $\bm{C^u}$, and of dimension $m$ in the case of $\bm{C^x}$. The
matrix $\bm{N^x}$ is square, of dimension $n-m$, and the various
rectangular blocks $\bm{0}$ are vanishing matrices whose dimensions are
clear from the context.  Dropping the hats ( $\widehat{ }$ ) for ease of
notation, the variable $v$ represents the set of normal variables $q$
and null variables $w$ of the characteristic problem in canonical form.
Additionally, $m$ functions $w_0\equiv (w_0^\nu(u,x^i))$ are given as
data on the surface ${\cal T}$ and $n-m$ functions $q_0\equiv
(q_0^\nu(x,x^i))$ are given as data on the surface ${\cal N}$. 

For the remainder of this section, we make the strong assumption that
$\bm{N^x}$ and $\bm{C}^i$ are symmetric. Multiplication of
(\ref{eq:wellstart}) by $v$ on the left then leads to a ``conservation
law'' of the form
\begin{equation}\label{eq:wellstep1}
	\partial_a(v\bm{C^a}v) + v\bm{R}v =0
\end{equation}

\noindent where $\bm{R} \equiv 2\bm{D} -\partial_a\bm{C^a}$. We now
integrate this conservation law in an appropriate volume $\cal V$ of
$\mathbb{R}^n$. Our volume is a ``hyperprism'' limited by the surface
$u=0$ from ``below'', the surface $x=0$ on the ``left'', and the surface
$u+x=T$, for an arbitrary constant $T$, on the ``top''. We assume there
are no boundaries in the remaining coordinate directions, in the sense
that the solutions $v$ will either be periodic functions of $x^i$ or
will decay sufficiently fast at large values of $x^i$ in order for the
integrals of their squares to exist.  The integration yields
\begin{subequations}\label{eq:wellstep2}
\begin{eqnarray}
&&\int_{\Sigma_T}v(\bm{C^u}+\bm{C^x})v \;d\Sigma_T
-\int_{\cal N}(v\bm{C^u}v) d{\cal N}-\nonumber\\
&&\hspace{1cm}-\int_{\cal T}(v\bm{C^x}v)\;d{\cal T} 
+\int_{\cal V}v\bm{R}v \;d{\cal V} =0.
\end{eqnarray}
\end{subequations}

\noindent  Clearly
\begin{equation}
\int_{\cal N}v\bm{C^u}v\; d{\cal N} 
=
\int_{\cal N} \sum^n_{\nu=m+1} (q_0^\nu)^2  d{\cal N}
\equiv ||q_0||^2,
\end{equation}

\noindent and
\begin{subequations}
\begin{eqnarray}
\int_{\cal T}v\bm{C^x}v \;d{\cal T} 
&=&
\int_{\cal T}q\bm{N^x}q \;d{\cal T}
+\int_{\cal T} \sum^m_{\nu=1} (w_0^\nu)^2  d{\cal T}\nonumber\\
&\equiv&\int_{\cal T}q\bm{N^x}q \;d{\cal T}
+ ||w_0||^2
\end{eqnarray}
\end{subequations}

Thus Eq.~(\ref{eq:wellstep2}) is rearranged to read
\begin{subequations}\label{eq:wellstep3}
\begin{eqnarray}
\int_{\Sigma_T}v(\bm{C^u}+\bm{C^x})v \;d\Sigma_T
&=&
||q_0||^2+||w_0||^2
+\int_{\cal T}q\bm{N^x}q \;d{\cal T}\nonumber\\
&&
-\int_{\cal V}v\bm{R}v \,d{\cal V}.
\end{eqnarray}
\end{subequations}

\noindent If $\bm{N^x}$ is non-positive definite but also such that
$\bm{1}+\bm{N^x}$ is positive definite, we can define the norm of the
solution $v$ on the surface $\Sigma_T$ by
\begin{equation}
||v||^2_T \equiv \int_{\Sigma_T}v(\bm{C^u}+\bm{C^x})v \;d\Sigma_T,
\end{equation}

\noindent and Eq.~(\ref{eq:wellstep3}) implies
\begin{equation}\label{eq:wellstep4}
||v||^2_T
\le
||q_0||^2+||w_0||^2
-\int_{\cal V}v\bm{R}v \,d{\cal V}.
\end{equation}

\noindent In special case of constant coefficients with no
undifferentiated terms, namely $\bm{R}=\bm{0}$, Eq.~(\ref{eq:wellstep4})
takes the form
\begin{equation}\label{eq:wellstep5}
||v||^2_T
\le
||q_0||^2+||w_0||^2 ,
\end{equation}

\noindent which represents an \textit{a priori} estimate of the solution
in terms of the free data. It implies that the ``size'' of the solution
is controlled by the ``size'' of the data. We may interpret it as a
statement of well-posedness of the characteristic problem.  Clearly the
estimate holds in the presence of non-constant coefficients and
undifferentiated terms as long as $\bm{R}$ is non-negative definite. 

An estimate can still be drawn in the presence of a negative definite
bounded $\bm{R}$, but it is weaker and holds only for small values of
$T$, as we show next. 

Since $\bm{R}$ is negative definite, then 
\begin{equation}
-v\bm{R}v \le r \sum_{\nu=1}^n (v^\nu)^2
\end{equation}

\noindent where $r = $max$(|R_{ij}|)$ in the volume ${\cal V}$, assuming
that such a number $r$ exists. On the other hand, since $\bm{C^u}+
\bm{C^x}$ is positive definite and symmetric then all its eigenvalues
are positive and we have
\begin{equation}
v(\bm{C^u}+ \bm{C^x})v \ge c \sum_{\nu=1}^n (v^\nu)^2
\end{equation}

\noindent with $c$ being the smallest eigenvalue of $\bm{C^u}+ \bm{C^x}$.
This implies 

\begin{equation}
-v\bm{R}v \le \frac{r}{c} 
v(\bm{C^u}+ \bm{C^x})v
\end{equation}

\noindent Thus 
\begin{equation}
-\int_{\cal V} v\bm{R}v d{\cal V}
\le \frac{r}{c} \int_0^T 
||v||_t^2\;dt
\end{equation}

\noindent where $||v||^2_t$ is the norm of the solution on the surface 
$\Sigma_t$ given by $u+x=t$ for fixed value of
$t<T$.
Thus the inequality (\ref{eq:wellstep4}) implies
\begin{equation}\label{eq:wellstep6}
||v||^2_T
\le
||q_0||^2+||w_0||^2
+
\frac{r}{c} \int_0^T 
||v||_t^2\;dt.
\end{equation}

\noindent Here $||q_0||^2$ and $||w_0||^2$ are the norms of the normal and
null variables with respect to the surfaces ${\cal N}$ and ${\cal T}$ both
bounded by the spatial surface at $u+x=T$.  For any value of $t\le T$ we
can write down the same inequality 
\begin{equation}\label{eq:wellstep7}
||v||^2_t
\le
\int_{{\cal N}_t} {\textstyle \sum} (q^\nu)^2 d{\cal N}_t
+\int_{{\cal T}_t} {\textstyle \sum} (w^\nu)^2 d{\cal T}_t
+
\frac{r}{c} \int_0^t 
||v||_{t'}^2\;dt'
\end{equation}

\noindent where ${\cal N}_t$ and ${\cal T}_t$ are the subsets of ${\cal
N}$ and ${\cal T}$ bounded by $\Sigma_t$, respectively. Since both
integrals indicated are less than the norms $||q_0||^2$ and $||w_0||^2$
respectively, this implies
\begin{equation}\label{eq:wellstep8}
||v||^2_t
\le
||q_0||^2+||w_0||^2
+
\frac{r}{c} \int_0^t 
||v||_{t'}^2\;dt'
\end{equation}

\noindent Using this inequality recursively into the right-hand side of
(\ref{eq:wellstep6}) we have
\begin{eqnarray}\label{eq:wellstep9}
||v||^2_T &\le& 
\left(1+\frac{rT}{c}+\frac{(rT)^2}{2c^2}+...+\frac{(rT)^j}{j!c^j}
\right)\times\nonumber\\
&& \times \bigg(||q_0||^2+||w_0||^2 + \nonumber\\
&& {}+ (r/c)^{j+1}\int_0^T\!\!\!dt_1\int_0^{t_1}\!\!\!dt_2...\int_0^{t_j}
\hspace{-0.4cm}||v||^2_{t_{j+1}} dt_{j+1}\bigg) \nonumber\\
&&
\end{eqnarray}

\noindent for any given non-negative integer $j$. In the limit for
$j\to\infty$ the sequence in the right-hand side converges if
$(rT/c)<1$, in which case we have
\begin{eqnarray}\label{eq:wellstep10}
||v||^2_T &\le&  e^{(r/c)T}\Big(
 ||q_0||^2+||w_0||^2       \Big)
\end{eqnarray}

\noindent This is our final estimate for the solution in terms of the 
free data on the surfaces ${\cal N}$ and ${\cal T}$. The estimate
involves an exponential factor essentially due to the presence of
undifferentiated terms. The exponential factor depends on the
properties of the system of equations (the principal matrices and the
undifferentiated terms), but not on the choice of data. This is
analogous to the \textit{a priori} estimates for Cauchy problems with
undifferentiated terms. As usual in such cases, the estimate is useless
for large $T$, and, in particular, our proof only guarantees the
estimate for $T<c/r$. Perhaps with greater care the estimate can be
extended to longer values of $T$. 

Because the \textit{a priori} estimates (\ref{eq:wellstep10}) are
independent of the choice of data, we can say that our characteristic
problem is well-posed. The conditions under which we are able to derive
\textit{a priori} estimates thus become our \textit{criteria for
well-posedness of linear homogeneous characteristic problems in
canonical form:}\newline
{\bf i)} The principal matrices $\bm{C^a}$ are symmetric.\newline
{\bf ii)} The normal block of the principal $x-$matrix, 
denoted $\bm{N^x}$, is non-positive 
definite but such that $\bm{1}+\bm{N^x}$ is positive definite (namely, 
$-\bm{1}< \bm{N^x}\le \bm{0}$).

There is, clearly, no obstacle in generalizing the construction 
slightly to the case where the characteristic problem is cast into
``almost'' canonical form, namely, the case when the principal matrices
are 
\begin{equation}
\bm{C^u}=\left(\begin{array}{cc}
	\bm{N^u}&\bm{0}\\
	\bm{0}&\bm{0}
	\end{array}
\right), 
\hspace{1cm}
\bm{C^x}=\left(\begin{array}{cc}
	\bm{N^x}&\bm{0}\\
	\bm{0}&\bm{1}
	\end{array}
\right) ,
\end{equation}

\noindent which corresponds to a strictly canonical form up to a
transformation of normal variables among themselves. In this case, the
criterion is\newline
{\bf i)} \textit{The principal matrices $\bm{C^a}$ are symmetric 
and }\newline
{\bf ii)} \textit{The normal block of the principal $u-$ matrix, denoted
$\bm{N^u}$, is positive definite. The normal block of the 
principal $x-$matrix, denoted $\bm{N^x}$, is non-positive 
definite but such that $\bm{N^u}+\bm{N^x}$ is positive definite 
(namely, $-\bm{N^u}< \bm{N^x}\le \bm{0}$).}\\

\noindent If {\bf i)} and {\bf ii)} hold for a linear homogeneous
characteristic problem in ``almost'' canonical form, then the problem is
well posed in the sense that there exist \textit{a  priori} estimates of
the kind $||v||^2_T \le e^{KT}(||q_0||^2+||w_0||^2)$, where $K$ is a
constant independent of the data. This inequality is sufficient to
establish the stability of the solutions under small variations of the
data. Notice that  $-\bm{N^u}< \bm{N^x}\le \bm{0}$ is equivalent to the
requirement that the surfaces given by $\phi(y^a)+\psi(y^a)=T$ with fixed
value of $T$ are spatial with respect to the hyperbolic operator
$\bm{A}^a$, which in turn means that they can be interpreted as the level
surfaces of a time function $t(y^a) \equiv \phi(y^a)+\psi(y^a)$.   

As an illustration, we can see that the first-order form of the wave
equation, Eqs.~(\ref{characwave}), is well posed. For 
Eqs.~(\ref{characwave}) we have
\begin{equation}
\bm{N^u}=\left(\begin{array}{ccc}
	2&0&0\\
	0&1&0\\
	0&0&1
	\end{array}
\right), 
\hspace{1cm}
\bm{N^x}=\left(\begin{array}{rcc}
	-1&0&0\\
	 0&0&0\\
	 0&0&0
	\end{array}
\right) ,
\end{equation}

\noindent and also 
\begin{equation}
\bm{C^y}=-\frac{1}{\sqrt{2}}
	\left(\begin{array}{cccc}
	0&1&0&0\\
	1&0&0&1\\
	0&0&0&0\\
	0&1&0&0
	\end{array}
\right), 
\hspace{0.5cm}
\bm{C^z}=-\frac{1}{\sqrt{2}}
	\left(\begin{array}{cccc}
	0&0&1&0\\
	0&0&0&0\\
	1&0&0&1\\
	0&0&1&0
	\end{array}
\right). 
\end{equation}

Therefore all the conditions are satisfied, and the estimates follow.
In this case, the estimates are of the form (\ref{eq:wellstep5}) and
hold for any chosen $T$ because all the principal matrices have
constant coefficients and there are no undifferentiated terms
($\bm{R}=\bm{0}$).

\section{Concluding remarks}
\label{sec:4}

Characteristic problems for hyperbolic equations are rarely discussed
in the literature.  In fact, prior to Balean's
work~\cite{baleanthesis,balean1,balean2}, practically nothing was
known about the characteristic problem of the simplest hyperbolic
equation, that is, the wave equation. Balean discussed how to derive
estimates for the solutions of the wave equation in its standard
second-order form.  Balean's estimates differ markedly from ours. The
estimates for general linear characteristic problems of the
first-order that we present here constitute a direct generalization
of the estimates that we recently derived for the particular case of
solutions of the characteristic problem of the wave equation as a
first-order system of PDE's~\cite{cmpwave}. 

The value of the generalization that we present here resides in the
formulation of algebraic criteria sufficient for the existence of the
\textit{a priori} estimates.  We demonstrate
elsewhere~\cite{estimate} that these criteria allow us to
formulate the characteristic problem of the linearized Einstein
equations in a form that is guaranteed to be well posed.

Several issues of interest remain wide open. First, given a general
characteristic problem that is well posed in the sense that we
introduce here, it is not at all clear as yet whether estimates of
the derivatives of the solution in terms of the derivatives of the
data would exist as well.  We have  succeeded in deriving estimates
for the derivatives in the particular case of the characteristic
problem of the wave equation~\cite{cmpwave}. However, the derivation
depends strongly on the particular form of the hyperbolic operator of
the wave equation, and its generalization to arbitrary characteristic
problems is far from straightforward, quite unfortunately. 

Secondly, a sufficient criterion to establish well posedness of a
characteristic problem is useful, but a necessary criterion would,
perhaps, be invaluable as a means to rule out unstable problems with
an eye towards numerical applications.

Thirdly, whether or not all well-posed hyperbolic problems admit
well-posed characteristic problems in our sense might well be  the
most intriguing open question at this time.

\begin{acknowledgments}

I am indebted to Roberto G\'{o}mez for numerous conversations. This work was
supported by the NSF under grants No. PHY-9803301, No. PHY-0070624 and No.
PHY-0244752 to Duquesne University.

\end{acknowledgments}


\end{document}